\documentclass[longauth]{aa}  

\usepackage{graphicx}
\usepackage{txfonts}
\usepackage{xspace}
\usepackage{color}
\usepackage{url}
\usepackage{hyperref}

\usepackage{newtxtext,newtxmath}
\usepackage{color}

\usepackage[normalem]{ulem}

\begin{document}

\title{Pulsar-wind-nebula-powered Galactic center X-ray filament G0.13-0.11\\ Proof of the synchrotron nature by IXPE}

\author{
Eugene Churazov \inst{\ref{in:MPA},\ref{in:IKI}}
\and Ildar Khabibullin \inst{\ref{in:USM},\ref{in:MPA},\ref{in:IKI}}
\and Thibault Barnouin \inst{\ref{in:Strasbourg2}} 
\and Niccol\`{o} Bucciantini 
\inst{\ref{in:INAF-Arcetri},\ref{in:UniFI},\ref{in:INFN-FI}} 
\and Enrico     Costa \inst{\ref{in:INAF-IAPS}} 
\and Laura      Di Gesu \inst{\ref{in:ASI}} 
\and Alessandro Di Marco \inst{\ref{in:INAF-IAPS}}
\and Riccardo Ferrazzoli \inst{\ref{in:INAF-IAPS}} 
\and William Forman \inst{\ref{in:CfA}} 
\and Philip Kaaret \inst{\ref{in:NASA-MSFC}} 
\and Dawoon E. Kim \inst{\ref{in:INAF-IAPS},\ref{in:LaSap},\ref{in:INFN-Roma2}}  
\and Jeffery J. Kolodziejczak \inst{\ref{in:NASA-MSFC}} 
\and Ralph Kraft \inst{\ref{in:CfA}} 
\and Fr\'{e}d\'{e}ric Marin \inst{\ref{in:Strasbourg2}} 
\and Giorgio Matt  \inst{\ref{in:UniRoma3}}  
\and Michela Negro \inst{\ref{in:LSU}} 
\and Roger W. Romani \inst{\ref{in:Stanford}}
\and Stefano Silvestri \inst{\ref{in:INFN-PI}}
\and Paolo Soffitta \inst{\ref{in:INAF-IAPS}} 
\and Rashid Sunyaev \inst{\ref{in:MPA},\ref{in:IKI}}
\and Jiri Svoboda \inst{\ref{in:CAS-ASU}}
\and Alexey Vikhlinin \inst{\ref{in:CfA}} 
\and Martin C. Weisskopf \inst{\ref{in:NASA-MSFC}} 
\and Fei Xie \inst{\ref{in:GSU},\ref{in:INAF-IAPS}}
\and Iv\'an Agudo \inst{\ref{in:CSIC-IAA}}
\and Lucio A. Antonelli \inst{\ref{in:INAF-OAR},\ref{in:ASI-SSDC}} 
\and Matteo Bachetti \inst{\ref{in:INAF-OAC}} 
\and Luca Baldini  \inst{\ref{in:INFN-PI},      \ref{in:UniPI}} 
\and Wayne H. Baumgartner  \inst{\ref{in:NASA-MSFC}} 
\and Ronaldo Bellazzini  \inst{\ref{in:INFN-PI}} 
\and Stefano Bianchi \inst{\ref{in:UniRoma3}}  
\and Stephen D. Bongiorno \inst{\ref{in:NASA-MSFC}} 
\and Raffaella Bonino  \inst{\ref{in:INFN-TO},\ref{in:UniTO}}
\and Alessandro Brez  \inst{\ref{in:INFN-PI}} 
\and Fiamma Capitanio \inst{\ref{in:INAF-IAPS}}
\and Simone Castellano \inst{\ref{in:INFN-PI}}  
\and Elisabetta Cavazzuti \inst{\ref{in:ASI}} 
\and Chien-Ting Chen \inst{\ref{in:USRA-MSFC}}
\and Stefano Ciprini \inst{\ref{in:INFN-Roma2},\ref{in:ASI-SSDC}}
\and Alessandra De Rosa \inst{\ref{in:INAF-IAPS}} 
\and Ettore     Del Monte \inst{\ref{in:INAF-IAPS}} 
\and Niccol\`{o} Di Lalla \inst{\ref{in:Stanford}}
\and Immacolata Donnarumma \inst{\ref{in:ASI}}
\and Victor Doroshenko \inst{\ref{in:Tub}}
\and Michal Dov\v{c}iak \inst{\ref{in:CAS-ASU}}
\and Steven     R. Ehlert \inst{\ref{in:NASA-MSFC}}  
\and Teruaki Enoto \inst{\ref{in:RIKEN}}
\and Yuri Evangelista \inst{\ref{in:INAF-IAPS}}
\and Sergio     Fabiani \inst{\ref{in:INAF-IAPS}}
\and Javier     A. Garc\'{i}a \inst{\ref{in:GoddardXray}}
\and Shuichi Gunji \inst{\ref{in:Yamagata}} 
\and Kiyoshi Hayashida \inst{\ref{in:Osaka}} 
\and Jeremy Heyl \inst{\ref{in:UBC}}
\and Wataru     Iwakiri \inst{\ref{in:Chiba}} 
\and Svetlana G. Jorstad \inst{\ref{in:BU},\ref{in:SPBU}} 
\and Vladimir Karas \inst{\ref{in:CAS-ASU}}
\and Fabian     Kislat \inst{\ref{in:UNH}} 
\and Takao      Kitaguchi  \inst{\ref{in:RIKEN}} 
\and Henric Krawczynski  \inst{\ref{in:WUStL}}
\and Fabio La Monaca \inst{\ref{in:INAF-IAPS},\ref{in:UniRoma2},\ref{in:LaSap}} 
\and Luca Latronico  \inst{\ref{in:INFN-TO}} 
\and Ioannis Liodakis \inst{\ref{in:NASA-MSFC}}
\and Simone     Maldera \inst{\ref{in:INFN-TO}}  
\and Alberto Manfreda \inst{\ref{INFN-NA}}
\and Andrea     Marinucci \inst{\ref{in:ASI}} 
\and Alan P. Marscher \inst{\ref{in:BU}} 
\and Herman L. Marshall \inst{\ref{in:MIT}}
\and Francesco  Massaro \inst{\ref{in:INFN-TO},\ref{in:UniTO}} 
\and Ikuyuki Mitsuishi \inst{\ref{in:Nagoya}} 
\and Tsunefumi  Mizuno \inst{\ref{in:Hiroshima}} 
\and Fabio      Muleri \inst{\ref{in:INAF-IAPS}} 
\and Chi-Yung Ng \inst{\ref{in:HKU}}
\and Stephen L. O'Dell \inst{\ref{in:NASA-MSFC}}  
\and Nicola     Omodei \inst{\ref{in:Stanford}}
\and Chiara     Oppedisano \inst{\ref{in:INFN-TO}}  
\and Alessandro Papitto \inst{\ref{in:INAF-OAR}}
\and George     G. Pavlov \inst{\ref{in:PSU}}
\and Abel Lawrence Peirson \inst{\ref{in:Stanford}}
\and Matteo     Perri \inst{\ref{in:ASI-SSDC},\ref{in:INAF-OAR}}
\and Melissa Pesce-Rollins \inst{\ref{in:INFN-PI}} 
\and Pierre-Olivier     Petrucci \inst{\ref{in:Grenoble}} 
\and Maura Pilia \inst{\ref{in:INAF-OAC}} 
\and Andrea     Possenti \inst{\ref{in:INAF-OAC}} 
\and Juri Poutanen \inst{\ref{in:UTU}}
\and Simonetta  Puccetti \inst{\ref{in:ASI-SSDC}}
\and Brian D. Ramsey \inst{\ref{in:NASA-MSFC}}  
\and John Rankin \inst{\ref{in:INAF-IAPS}} 
\and Ajay Ratheesh \inst{\ref{in:INAF-IAPS}} 
\and Oliver     J. Roberts \inst{\ref{in:USRA-MSFC}}
\and Carmelo Sgr\`o \inst{\ref{in:INFN-PI}}  
\and Patrick Slane \inst{\ref{in:CfA}}  
\and Gloria     Spandre \inst{\ref{in:INFN-PI}} 
\and Douglas A. Swartz \inst{\ref{in:USRA-MSFC}}
\and Toru Tamagawa \inst{\ref{in:RIKEN}}
\and Fabrizio Tavecchio \inst{\ref{in:INAF-OAB}}
\and Roberto Taverna \inst{\ref{in:UniPD}} 
\and Yuzuru     Tawara \inst{\ref{in:Nagoya}}
\and Allyn F. Tennant \inst{\ref{in:NASA-MSFC}}  
\and Nicholas E. Thomas \inst{\ref{in:NASA-MSFC}}  
\and Francesco  Tombesi  \inst{\ref{in:UniRoma2},\ref{in:INFN-Roma2},\ref{in:UMd}}
\and Alessio Trois \inst{\ref{in:INAF-OAC}}
\and Sergey S. Tsygankov \inst{\ref{in:UTU}}
\and Roberto Turolla \inst{\ref{in:UniPD},\ref{in:MSSL}}
\and Jacco Vink \inst{\ref{in:Amsterdam}}
\and Kinwah     Wu \inst{\ref{in:MSSL}}
\and Silvia Zane  \inst{\ref{in:MSSL}}
          }
          
\institute{
Max Planck Institute for Astrophysics, Karl-Schwarzschild-Str. 1, D-85741 Garching, Germany \label{in:MPA}
\email{churazov@mpa-garching.mpg.de}
\and 
Space Research Institute (IKI), Profsoyuznaya 84/32, Moscow 117997, Russia\label{in:IKI}
\and
Universitäts-Sternwarte, Fakultät für Physik, Ludwig-Maximilians-Universität München, Scheinerstr.1, 81679 München, Germany\label{in:USM}
\and 
Universit\'{e} de Strasbourg, CNRS, Observatoire Astronomique de Strasbourg, UMR 7550, 67000 Strasbourg, France \label{in:Strasbourg2}
\and   
INAF Osservatorio Astrofisico di Arcetri, Largo Enrico Fermi 5, 50125 Firenze, Italy 
\label{in:INAF-Arcetri} 
\and  
Dipartimento di Fisica e Astronomia, Universit\`{a} degli Studi di Firenze, Via Sansone 1, 50019 Sesto Fiorentino (FI), Italy \label{in:UniFI} 
\and   
Istituto Nazionale di Fisica Nucleare, Sezione di Firenze, Via Sansone 1, 50019 Sesto Fiorentino (FI), Italy \label{in:INFN-FI}
\and 
INAF Istituto di Astrofisica e Planetologia Spaziali, Via del Fosso del Cavaliere 100, 00133 Roma, Italy \label{in:INAF-IAPS}
\and 
Agenzia Spaziale Italiana, Via del Politecnico snc, 00133 Roma, Italy \label{in:ASI}
\and 
Center for Astrophysics, Harvard \& Smithsonian, 60 Garden St, Cambridge, MA 02138, USA \label{in:CfA} 
\and 
NASA Marshall Space Flight Center, Huntsville, AL 35812, USA \label{in:NASA-MSFC}
\and 
Dipartimento di Fisica, Università degli Studi di Roma "La
Sapienza", Piazzale Aldo Moro 5, 00185 Roma, Italy \label{in:LaSap}
\and 
Istituto Nazionale di Fisica Nucleare, Sezione di Roma ``Tor Vergata'', Via della Ricerca Scientifica 1, 00133 Roma, Italy 
\label{in:INFN-Roma2}
%\and 
%Universit\'{e} de Strasbourg, CNRS, Observatoire Astronomique de Strasbourg, UMR 7550, 67000 Strasbourg, France \label{in:Strasbourg}
\and 
Dipartimento di Matematica e Fisica, Universit\`a degli Studi Roma Tre, via della Vasca Navale 84, 00146 Roma, Italy  \label{in:UniRoma3}
\and 
Department of Physics and Astronomy, Louisiana State University, Baton Rouge, LA 70803, USA \label{in:LSU}
\and 
Department of Physics and Kavli Institute for Particle Astrophysics and Cosmology, Stanford University, Stanford, California 94305, USA  \label{in:Stanford}
\and 
Istituto Nazionale di Fisica Nucleare, Sezione di Pisa, Largo B. Pontecorvo 3, 56127 Pisa, Italy \label{in:INFN-PI}
\and 
Astronomical Institute of the Czech Academy of Sciences, Bo\v{c}n\'{i} II 1401/1, 14100 Praha 4, Czech Republic \label{in:CAS-ASU}
\and 
Guangxi Key Laboratory for Relativistic Astrophysics, School of Physical Science and Technology, Guangxi University, Nanning 530004, China \label{in:GSU}
\and 
Instituto de Astrof\'{i}sica de Andaluc\'{i}a -- CSIC, Glorieta de la Astronom\'{i}a s/n, 18008 Granada, Spain \label{in:CSIC-IAA}
\and 
INAF Osservatorio Astronomico di Roma, Via Frascati 33, 00040 Monte Porzio Catone (RM), Italy \label{in:INAF-OAR} 
\and 
Space Science Data Center, Agenzia Spaziale Italiana, Via del Politecnico snc, 00133 Roma, Italy \label{in:ASI-SSDC}
 \and
INAF Osservatorio Astronomico di Cagliari, Via della Scienza 5, 09047 Selargius (CA), Italy  \label{in:INAF-OAC}
\and  
Dipartimento di Fisica, Universit\`{a} di Pisa, Largo B. Pontecorvo 3, 56127 Pisa, Italy \label{in:UniPI} 
\and  
Istituto Nazionale di Fisica Nucleare, Sezione di Torino, Via Pietro Giuria 1, 10125 Torino, Italy  \label{in:INFN-TO}      
\and  
Dipartimento di Fisica, Universit\`{a} degli Studi di Torino, Via Pietro Giuria 1, 10125 Torino, Italy \label{in:UniTO} 
\and 
Science and Technology Institute, Universities Space Research Association, Huntsville, AL 35805, USA \label{in:USRA-MSFC}
\and
Institut f\"ur Astronomie und Astrophysik, Universit\"at T\"ubingen, Sand 1, D-72076 T\"ubingen, Germany \label{in:Tub}
\and 
RIKEN Cluster for Pioneering Research, 2-1 Hirosawa, Wako, Saitama 351-0198, Japan \label{in:RIKEN}
\and 
X-ray Astrophysics Laboratory, NASA Goddard Space Flight Center, Greenbelt, MD 20771, USA \label{in:GoddardXray}
\and 
Yamagata University,1-4-12 Kojirakawa-machi, Yamagata-shi 990-8560, Japan \label{in:Yamagata}
\and 
Osaka University, 1-1 Yamadaoka, Suita, Osaka 565-0871, Japan \label{in:Osaka}
\and 
University of British Columbia, Vancouver, BC V6T 1Z4, Canada \label{in:UBC}
\and 
International Center for Hadron Astrophysics, Chiba University, Chiba 263-8522, Japan \label{in:Chiba}
\and
Institute for Astrophysical Research, Boston University, 725 Commonwealth Avenue, Boston, MA 02215, USA \label{in:BU} 
\and 
Department of Astrophysics, St. Petersburg State University, Universitetsky pr. 28, Petrodvoretz, 198504 St. Petersburg, Russia \label{in:SPBU} 
\and 
Department of Physics and Astronomy and Space Science Center, University of New Hampshire, Durham, NH 03824, USA \label{in:UNH} 
\and 
Physics Department and McDonnell Center for the Space Sciences, Washington University in St. Louis, St. Louis, MO 63130, USA \label{in:WUStL}
\and 
Istituto Nazionale di Fisica Nucleare, Sezione di Napoli, Strada Comunale Cinthia, 80126 Napoli, Italy \label{INFN-NA}
\and 
MIT Kavli Institute for Astrophysics and Space Research, Massachusetts Institute of Technology, 77 Massachusetts Avenue, Cambridge, MA 02139, USA \label{in:MIT}
\and 
Graduate School of Science, Division of Particle and Astrophysical Science, Nagoya University, Furo-cho, Chikusa-ku, Nagoya, Aichi 464-8602, Japan \label{in:Nagoya}
\and 
Hiroshima Astrophysical Science Center, Hiroshima University, 1-3-1 Kagamiyama, Higashi-Hiroshima, Hiroshima 739-8526, Japan \label{in:Hiroshima}
\and 
Department of Physics, University of Hong Kong, Pokfulam, Hong Kong \label{in:HKU}
\and 
Department of Astronomy and Astrophysics, Pennsylvania State University, University Park, PA 16801, USA \label{in:PSU}
\and 
Universit\'{e} Grenoble Alpes, CNRS, IPAG, 38000 Grenoble, France \label{in:Grenoble}
\and Department of Physics and Astronomy, FI-20014 University of Turku,  Finland \label{in:UTU} \\ 
\and 
INAF Osservatorio Astronomico di Brera, via E. Bianchi 46, 23807 Merate (LC), Italy \label{in:INAF-OAB}
\and 
Dipartimento di Fisica e Astronomia, Universit\`{a} degli Studi di Padova, Via Marzolo 8, 35131 Padova, Italy \label{in:UniPD}
\and
Dipartimento di Fisica, Universit\`{a} degli Studi di Roma ``Tor Vergata'', Via della Ricerca Scientifica 1, 00133 Roma, Italy \label{in:UniRoma2}
\and
Department of Astronomy, University of Maryland, College Park, Maryland 20742, USA \label{in:UMd}
\and 
Mullard Space Science Laboratory, University College London, Holmbury St Mary, Dorking, Surrey RH5 6NT, UK \label{in:MSSL}
\and 
Anton Pannekoek Institute for Astronomy \& GRAPPA, University of Amsterdam, Science Park 904, 1098 XH Amsterdam, The Netherlands  \label{in:Amsterdam}
}

\titlerunning{Synchrotron nature of PWN X-ray emission}
\authorrunning{Churazov et al.}

\date{2023}

%\shorttitle{Synchrotron nature of PWN X-Ray emission}
%\shortauthors{Churazov et al.}

% The list of authors, and the short list which is used in the headers.
% If you need two or more lines of authors, add an extra line using \newauthor

%\author[Churazov et al.]{E.~Churazov,$^{1,2}$ I.~Khabibullin,$^{3,1,2}$, T.Barnouin,  N.Bucciantini, E.Costa, L.Di Gesu,  A.Di Marco, 
%\and 
%%J.Svoboda, A.Vikhlinin, F.Xie, ...
%\\
%\\
%and others $^{2}$
% List of institutions
%$^1$~Max Planck Institute for Astrophysics, Karl-Schwarzschild-Str. 1, D-85741 Garching, Germany  \\
%$^2$~Space Research Institute (IKI), Profsoyuznaya 84/32, Moscow 117997, Russia \\
%$^3$~Universitäts-Sternwarte, Fakultät für Physik, Ludwig-Maximilians-Universität München, Scheinerstr.1, 81679 München, Germany\\
%}
%\author[Authors]{Eugene Churazov, Ildar Khabibullin, Rashid Sunyaev, Alexey Vikhlinin
%,$^{1,2}$
%\author[E. Churazov et al.]{E. Churazov,$^{1,2}$\thanks{E-mail:churazov@mpa-garching.mpg.de} I. Khabibullin,$^{1,2}$ R. Sunyaev$^{1,2}$
%and others $^{2}$
%\\
% List of institutions
%$^1$~Max Planck Institute for Astrophysics, Karl-Schwarzschild-Str. 1, D-85741 Garching, Germany  \\
%$^2$~Space Research Institute (IKI), Profsoyuznaya 84/32, Moscow 117997, Russia 
%}

% These dates will be filled out by the publisher
%\date{Accepted XXX. Received YYY; in original form ZZZ}

% Enter the current year, for the copyright statements etc.
%\pubyear{2015}

% Don't change these lines
%\label{firstpage}
%\pagerange{\pageref{firstpage}--\pageref{lastpage}}
%\maketitle

% Abstract of the paper
\abstract{
We report the discovery of X-ray polarization from the X-ray-bright filament.  G0.13-0.11 in the Galactic center (GC) region. This filament features a bright, hard X-ray source that is most plausibly a pulsar wind nebula (PWN) and an extended and structured diffuse component. Combining the polarization signal from IXPE with the imaging/spectroscopic data from \textit{Chandra}, we find that X-ray emission of G0.13-0.11 is highly polarized PD=$57(\pm18)$\% in the 3-6 keV band, while the polarization angle is PA=$21^\circ(\pm9^\circ)$. This high degree of polarization proves the synchrotron origin of the X-ray emission from G0.13-0.11. In turn, the measured polarization angle implies that the X-ray emission is polarized approximately perpendicular to a sequence of nonthermal radio filaments that may be part of the GC Radio Arc. The magnetic field on the order of $100\,{\rm\mu G}$ appears to be preferentially ordered along the filaments. The above field strength is the fiducial value that makes our model self-consistent, while the other conclusions are largely model independent.}

% Select between one and six entries from the list of approved keywords.
% Don't make up new ones.
%\begin{keywords}
% stars: pulsars; Astrophysics - High Energy Astrophysical Phenomena
%\end{keywords}

%%%%%%%%%%%%%%%%%%%%%%%%%%%%%%%%%%%%%%%%%%%%%%%%%%

%%%%%%%%%%%%%%%%% BODY OF PAPER %%%%%%%%%%%%%%%%%%

%\section{Figures}

%See six figures illustrating the main findings below.

\maketitle

\section{Introduction}
The origin of spectacular radio filaments (including the most prominent Radio Arc) in the Galactic center (GC) region and the role of pulsar wind nebulae (PWNe) in their appearance is an actively debated topic \citep[e.g.,][]{2004ApJS..155..421Y}. Many of the observed radio filaments have spectral and polarization properties suggesting that they are dominated by non-thermal emission \citep[e.g][]{1984Natur.310..557Y,1986AJ.....92..818T,1999ApJ...526..727L,2004ApJ...607..302L,2019Natur.573..235H}. Hereafter, we refer to them as nonthermal filaments (NTFs).   

%\ec{ADD HERE}
%\ec{PWN reviews, Gansler \& Slane, Kargaltsev, Bykov, Olmi}
%\ec{G0.13-0.11 in X-rays Wang+2002, Zhang+2020}

Some of the NTFs have known X-ray counterparts. Here, we focus on one of them -- an X-ray/radio thread  G0.13-0.11. 
In the X-ray band, G0.13-0.11 was studied by \citealt[][]{2002ApJ...568L.121Y,2002ApJ...581.1148W}, and \citealt[][]{2020ApJ...893....3Z} using \textit{Chandra} and NuSTAR data. The case of  G0.13-0.11 is especially interesting due to the close correspondence of X-ray structures and a subsystem of the Radio Arc filaments noted by \cite{2002ApJ...568L.121Y,2002ApJ...581.1148W}.   Based on a filamentary X-ray appearance, the presence of a compact source, and a nonthermal spectrum, the latter two studies focused on the interpretation of G0.13-0.11 as a PWN with a filament formed by leptons escaping from the supersonically moving PWN into the ambient magnetic field \citep[see, e.g.,][for reviews]{2006ARA&A..44...17G,2017JPlPh..83e6301K,2017SSRv..207..235B,2023PASA...40....7O}. 

In the standard PWN scenario, described in the aforementioned reviews, a pulsar generates an ultrarelativistic electron-positron wind that passes through its termination shock and gives rise to a wind nebula filled with high-energy leptons. For a supersonically moving PWN, a bow shock is formed ahead of the source.
In this class of models, known as bow-shock PWN (BSPWN), the leptons with a sufficiently large Lorentz factor (and, hence, a large gyroradius)  can cross the bow-shock region and escape the source \citep{Bandiera08a}.   \citealt{2017SSRv..207..235B} (see their Section 6) discussed conditions needed to produce very hard spectra (dominated by the highest energies) and a reconnection scenario that can lead to asymmetric or one-sided X-ray structures observed in several BSPWNs, such as the Lighthouse \citep{2011A&A...533A..74P} or Guitar \citep{2022ApJ...939...70D} nebulae. In this scenario, the mutual orientation of the PWN and the ambient magnetic fields set the topology of the field lines in a way reminiscent of Earth's magnetosphere embedded in the solar wind. Numerical magneto-hydrodynamical (MHD) models of these objects appear to be in reasonable agreement with observations \citep[e.g.,][]{2019MNRAS.484.4760B,2019MNRAS.490.3608O}. In these models, the radio and X-ray fluxes from threads are produced by synchrotron emission of relativistic leptons and hence should be polarized, if the magnetic field is at least partly ordered. Now, IXPE \citep{2022JATIS...8b6002W} can test this prediction in the X-ray regime. In the X-ray regime, the polarization angle (PA) is expected to be perpendicular to the orientation of the magnetic field, and, unlike the radio data \citep[e.g.,][]{2019ApJ...884..170P,2021ApJ...923...82P}, it is not affected by the Faraday rotation and depolarization due to foreground magnetized plasma. 

Should IXPE observations place tight upper limits on the polarization degree, this would imply either that the diffuse X-ray emission is not due to the synchrotron mechanism or that the magnetic field is highly disordered. The former assumption would contradict the basic model of PWN-powered X-ray extended structures and create a major problem not only for  G0.13-0.11 but for the entire class of these objects. The latter possibility would imply that magnetic field topology in the regions contributing to the X-ray emission is very different (much more tangled) from that in the nearby radio-emitting filaments, which have highly ordered fields based on observations at 1-100 GHz frequencies \cite[e.g.,][]{2021ApJ...920....6G}. Yet another possibility, discussed in \cite{2002ApJ...568L.121Y}, is that G0.13-0.11 is not a PWN, and low-energy cosmic-ray electrons produce the power-law continuum by bremsstrahlung emission in the X-ray band.

%\citep{1984Natur.310..557Y}
%spectral index 6-20 cm between -0.2 and 0.2
%mostly unpolarized, except for two locations where it is perp to fil. 
%pol 15-25%, Faradya rotation?

%\citep{2019Natur.573..235H}
% 1284 MHz, 6"
% cooling time 1-2 Myr

%\cite[][]{2002ApJ...581.1148W}
% X-ray age 1.3 yr /E_kev^{0.5}/B_mG^{1.5} ~ agrees with my estimates (d-check)
% 40" size at c is crossed in 5 yr. they assume c/sqrt(3) and said that B<0.3mG

% Reich, Sofue & Matsuo 2000 = high-freq. slightly shifted from 20 cm

%X-ray emission co-spatial with NTFs

%The case of G0.13-0.11 

%\cite[][]{2002ApJ...581.1148W,2020ApJ...893....3Z}

%------------------------
\begin{figure*}
\centering
\includegraphics[angle=0,clip,trim=1.8cm 10.3cm 3cm 8.5cm,width=2.1\columnwidth]{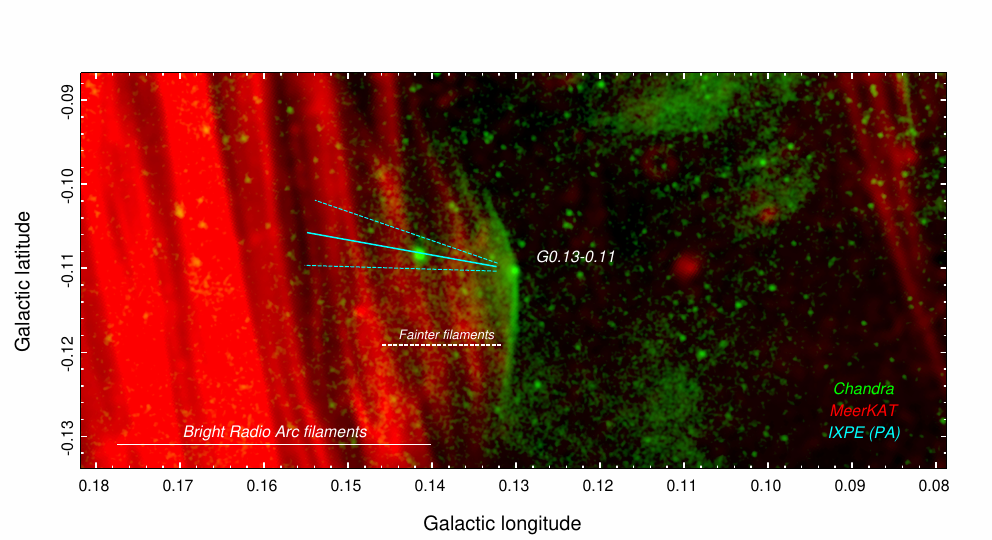}
\caption{Combination of radio (MeerKAT; 1.3~GHz, red) and X-ray (\textit{Chandra}; 2.3-8 keV, Log scale, smoothed with $\sigma=0.5''$ Gaussian, green)  images of the G0.13-0.11 region in Galactic coordinates. The cyan lines show the
constraints on the  
polarization angle (PA) of the  G0.13-0.11 emission derived from the IXPE data 
(best-fitting value $\pm 1\sigma$ error; i.e., $21\pm 9$ degrees).
%IXPE polarization is shown as cyan lines while the IXPE image is shown in Fig. 2.
The prominent red vertical threads on the left side of the image belong to the NTFs of the Radio Arc. G0.13-0.11 is the green feature to the east of a sequence of fainter NTFs near the center of the image. G0.13-0.11 consists of (i) a compact core (the bright green dot at the center of the figure; presumably the pulsar itself and its PWN); (ii) a pair of bright "wings"; and (iii) a more diffuse (possibly structured) X-ray glow to the left of the "wings" (at higher Galactic longitude). 
%The cyan lines show the Polarization Angle (PA) of the  G0.13-0.11 emission (best-fitting value $\pm 1\sigma$ error, i.e. $21\pm 9$ degrees) measured by IXPE. 
The polarization direction is approximately perpendicular to the X-ray "wings" and the nearest NTFs. Coupled with the large polarization degree, this proves that X-rays are produced by synchrotron emission of $\sim$ TeV electrons.  
}
\label{fig:outlook}
\end{figure*}
%-------------------------
%(i) a compact source _containing the pulsar and the actual bow shock_ that presumably provides

%266.510000     -28.890000
%266.570000     -28.890000
%3.15197'

\section{Data}
In this work, we used the data of two IXPE observations of the GC region, more specifically, a complex of molecular clouds $\sim 0.1$ degrees to the east of Sgr~A* \citep{2023Natur.619...41M}. These observations with a total clean exposure time of $1.8$~Msec,  were performed in two parts - in February 2022 and September 2023. These two observations were centered on (RA,Dec) (266.51,-28.89) and (266.57,-28.89), respectively, with an offset between them of $\sim 3'$.
The data were processed with the standard IXPE pipeline. 
The output FITS files of this pipeline contain the event-by-event Stokes parameters \citep[see][]{Kislat2015} from which the polarization observables of the X-ray radiation can be derived. The data products are publicly available for use by the international astrophysics community at the High-Energy Astrophysics Science Archive Research Center (HEASARC, at the NASA Goddard Space Flight Center). An energy-dependent particle background rejection algorithm based on photoelectron-track ellipticity was applied to these level-2 event files; it allows the removal of $\sim40\%$ of the instrumental background \citep{dimarco23}. In addition, observation times affected by increased background due to solar activity were removed 
%\citep[see][for additional details on the procedure]{2023Natur.619...41M}  
(see \citealt{2023Natur.619...41M} for additional details on the procedure).

High spatial resolution X-ray images were obtained with \textit{Chandra} over multiple observing campaigns from 2000-2022. The total exposure of the \textit{Chandra} pointings with G0.13--0.11 located in the central region of the field of view (FOV) with high spatial resolution is 1.4~Msec. The \textit{Chandra} data reduction follows a standard procedure based on the latest versions of the data reduction software (CIAO v.\ 4.14) and calibration (CALDB v.\ 4.9.8). Our particular approach and analysis steps are described in detail in \cite{Vikhlinin2009}. Briefly, they include the identification and removal of high background periods, the correction of photon energies for the time and detector temperature dependence of the charge transfer inefficiency and gain, and the creation of matching background datasets using blank sky observations with exposure times similar to the GC pointings. For the analysis presented here, we used the combined flat-fielded and background-subtracted \textit{Chandra} image in the 2--8~keV band and spectra extracted in several elliptical regions, as described below. Following the standard approach for analyzing \textit{Chandra} spectra of extended sources, we have generated the spectral response files that combine the position-dependent ACIS calibration with the weights proportional to the observed brightness.
In the radio band, we used publicly available MeerKAT images\footnote{\url{https://doi.org/10.48479/fyst-hj47}} at the central frequency of $1.3$~GHz and $\sim 6''$ angular resolution (see \citealt{2022ApJ...925..165H} for a detailed description).
%\citep[see][\LEt{please remove comma***}for a detailed description]{2022ApJ...925..165H}.

Figure~\ref{fig:outlook} shows a combination of the \textit{Chandra} X-ray image in the 2.3-8 keV band (in green) and the MeerKAT radio image (in red). G0.13-0.11 is a green feature to the right of a triangular-shaped sequence of NTFs near the center of the image. The cyan lines show the polarization angle of the G0.13-0.11 X-ray emission (best-fitting value $\pm 1\sigma$ error) measured by IXPE.

\section{X-ray images and spectra} 
\begin{figure*}
\centering
\includegraphics[angle=0,trim=1.5cm 12cm 2.5cm 10cm,clip,width=2.06\columnwidth]{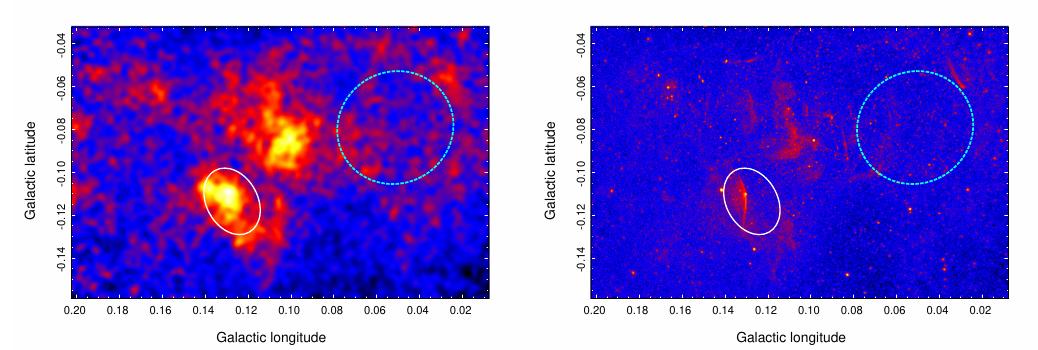}
\caption{3-6 keV IXPE (left panel) and \textit{Chandra} (right panel) X-ray images. Two elliptical regions were used for spectra extraction. The white ellipse covers the G0.13-0.11 area. Given the uncertainties with the IXPE astrometry, a relatively large region was selected guided by the IXPE image. The bigger cyan region was used as a background region for spectral analysis. The \textit{Chandra} image is in logarithmic scale to show more clearly bright point sources and faint diffuse emission. For the IXPE image, a linear scale is used.
}
\label{fig:ximages}
\end{figure*}

The X-ray images of the G0.13-0.11 region in the 2-8 keV band obtained by IXPE and \textit{Chandra} are shown in Fig.~\ref{fig:ximages}. The images were smoothed with a Gaussian filter with $\sigma=5''$ and $0.5''$ for IXPE and \textit{Chandra}, respectively.  G0.13-0.11 is located in the bottom left corner of the image and is clearly visible as a pair of bright wings in the \textit{Chandra} image. Sgr~A* is outside of the region shown, $\sim 0.06$ degrees from the right boundary.

%------------------------
%-------------------------

%------------------------
\begin{figure}
\centering
\includegraphics[angle=0,clip,trim=2.5cm 7.5cm 5cm 6.5cm,width=0.99\columnwidth]{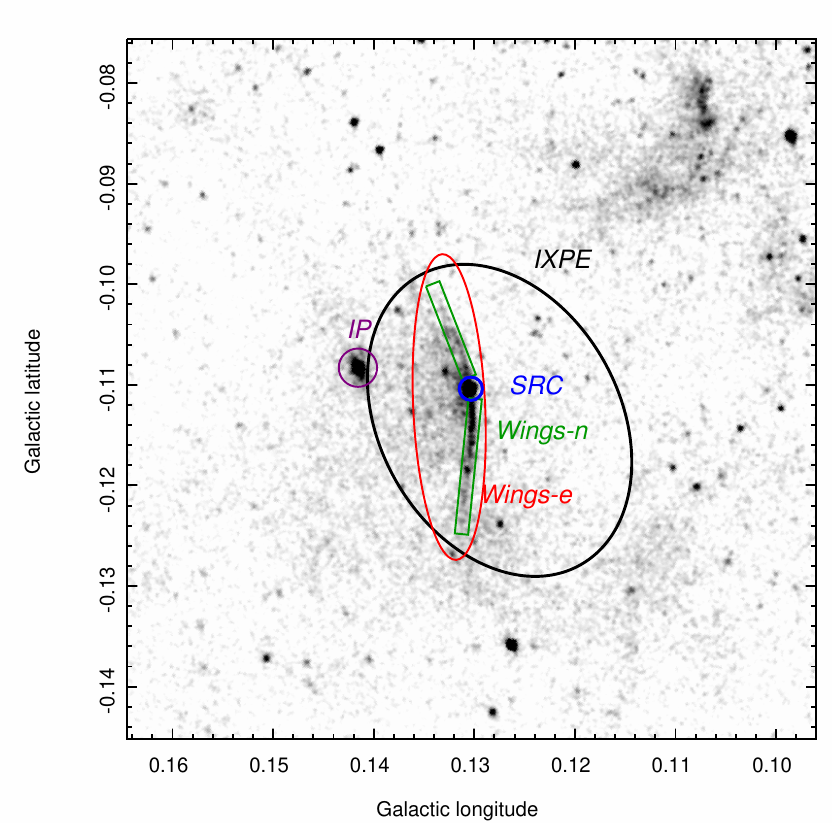}
\caption{Regions used for spectral extraction from 
\textit{Chandra} data. The black ellipse called IXPE is the same as the source region in Fig.~\ref{fig:ximages}. The other three regions (SRC, Wings-n, and Wings-e) cover the compact source and the brighter parts of the X-ray bow. The SRC and Wings-n regions do not overlap. Larger regions (Wings-e and IXPE) include all the \textit{Chandra} counts inside them; i.e., the IXPE region includes the counts from SRC, Wings-n, and almost the entire Wings-e counts. The purple circle labeled as IP shows the extraction region for the intermediate polar CXOUGC J174622.7–285218.
}
\label{fig:chandra_regs}
\end{figure}
%-------------------------

\begin{figure}
\centering
\includegraphics[angle=0,clip,trim=1cm 5.5cm 1cm 3cm,width=0.99\columnwidth]{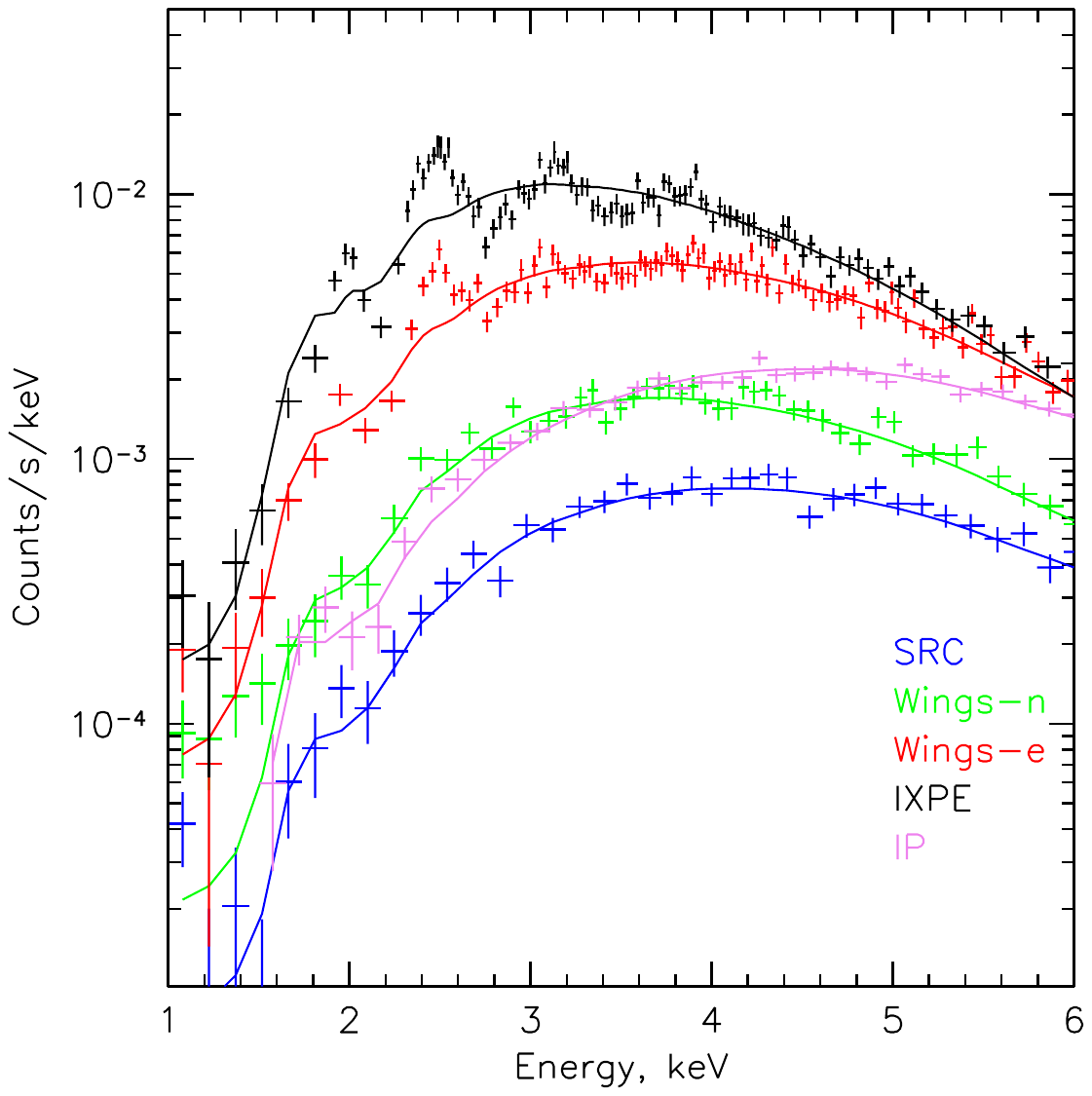}
\caption{Background-subtracted \textit{Chandra} spectra in the vicinity of G0.13-0.11. The colors correspond to the regions shown in Fig.~\ref{fig:chandra_regs} and are identified in this figure. The solid lines show the best-fitting absorbed power-law model fits to these spectra. While a pure power law can describe the compact source spectrum (blue) reasonably well, other regions clearly contain a non-negligible contribution of thermal emission that gives rise to emission lines at low energies. The spectrum of the intermediate polar CXOUGC J174622.7–285218 is shown in purple.  
}
\label{fig:spec_chandra}
\end{figure}

With the approach outlined above, IXPE will provide the polarized emission (Q and U spectra) of the entire region, while \textit{Chandra} data resolve the spectrum into spatially distinct components, in particular, the nonthermal component.  The \textit{Chandra} image with the additional regions used for spectra extraction is shown in Fig.~\ref{fig:chandra_regs}, including a point source CXOGCS~J174621.5-285256 marked with the blue circle (SRC), two narrow green boxes covering the brightest part of the box-shaped structure (Wings-n), a larger red ellipse that covers the "wings" and a more extended structure to the left of the wings (Wings-e), and the entire IXPE region (black). The latter region ($\approx 43\times60''$) is the same as shown in Fig.~\ref{fig:ximages}. In the 2022 and 2023 IXPE observations, the IXPE region was $\sim 4$ and $\sim 1.8$ arcminutes from the center of the instrument FoV, respectively. We note that while the IXPE instrumental half-power diameter (HPD) is ~25", flexure and aspect jitter always somewhat
blur the images. While this can be largely removed during typical IXPE observations of
bright point sources, in fields with a faint extended emission aspect, correction is more difficult. 
Despite extra care being taken in aligning the images, such blur remains, and
so we adopted generous extraction apertures. This should help to collect most of the photons from the polarized source and make the resulting spectrum less sensitive to the exact positioning of the extraction region. 
The spectra extracted from the \textit{Chandra} data (corrected for the background extracted from the almost circular region ($\approx 100\times94''$) shown with the dashed line in  Fig.~\ref{fig:ximages})  are shown in Fig.~\ref{fig:spec_chandra}. The diffuse X-ray emission of the GC region consists of several spectral components, some of which are highly variable in space and time and contain bright emission lines such as the lines at 6.4 and 6.7 keV of neutral and He-like iron. The former line is due to the reflection of X-rays by the neutral/molecular medium \citep[e.g.,][]{1980SvAL....6..353V,1996PASJ...48..249K}, while the latter is a combination of numerous accreting stellar-mass objects that produce "thermal" spectra \citep[e.g.,][]{2009Natur.458.1142R}. This means that there is no well-defined and "optimal" choice of the background region. Since in this study we are interested in the nonthermal component, the background region was selected to have a comparable mix of the reflection and thermal spectra. No fine-tuning was done when selecting the background region, although we verified that other reasonable choices do not significantly affect our final results. To mitigate the uncertainties associated with the contributions of the reflection component above 6 keV and strong lines below 3 keV coupled with strong low-energy photoelectric absorption, we restricted the analysis of IXPE data to the 3-6 keV band. The higher energy resolution of \textit{Chandra} permits the spectral analysis in a broader band, illustrating strong suppression of X-ray flux below 2-3 keV due to photo-electric absorption with a typical column density $N_{\rm H}\sim 8\,10^{22}\,{\rm cm^{-2}}$ (see Table~\ref{tab:spec}). 

Simple absorbed power-law model fits to the spectra are shown with solid lines. Clearly, strong emission lines (e.g., of strongly ionized Si or S) are present in the \textit{Chandra} spectrum of the "IXPE" region even after subtracting the spectrum extracted from the background region. Somewhat weaker lines are still seen in the Wings-e spectrum, while the Wings-n and SRC spectra seem to be reasonably well described by the power-law model.

%------------------------
%-------------------------

\begin{table*}
%\centering
\caption{Absorbed power-law model fits to \textit{Chandra} spectra for three regions of G0.11-0.13 (see Fig.~\ref{fig:chandra_regs} for region definition). The observed 2-10~keV fluxes (i.e., including the effects of the low-energy photoelectric absorption) are quoted. The statistical uncertainties in the model parameters $N_{\rm H}$ and $\Gamma$ are given.}
\vspace{0.1cm}
\begin{center}
\begin{tabular}{lrrrrr}
\hline
Region & $N_{\rm H}$   & $\Gamma$ & $F_{2-10\,{\rm keV}}$                       & Area       & $\chi^2$ (d.o.f.)  \\ 
       & $10^{22}\,\rm cm^{-2}$ &          &$ {\rm erg \,s^{-1}\,cm^{-2}}$ & arcsec$^2$ &  \\
\hline
\hline
"SRC" & $7.7\pm 0.6$ & $1.35\pm 0.14$ & $1.1\,10^{-13}$ & $54$ & 465 (478) \\ 
"Wings-n" & $7.6\pm 0.4$ & $2.27\pm 0.12$ & $1.6\,10^{-13}$ & 420 & 506 (478) \\ 
"Wings-e" & $7.1\pm 0.2$ & $2.40\pm 0.07$ & $5.0\,10^{-13}$ & 2213 & 645 (478) \\ 
Entire "IXPE" region & $8.1\pm 0.2$ & $3.75\pm 0.07$ & $6.4\,10^{-13}$ & 8000 & 1010 (478) \\ 
\hline
\hline
\end{tabular}
\end{center}
\label{tab:spec}
\end{table*}

\begin{figure}
\centering
\includegraphics[angle=0,trim=0.5cm 5.5cm 1cm 3cm,width=0.99\columnwidth]{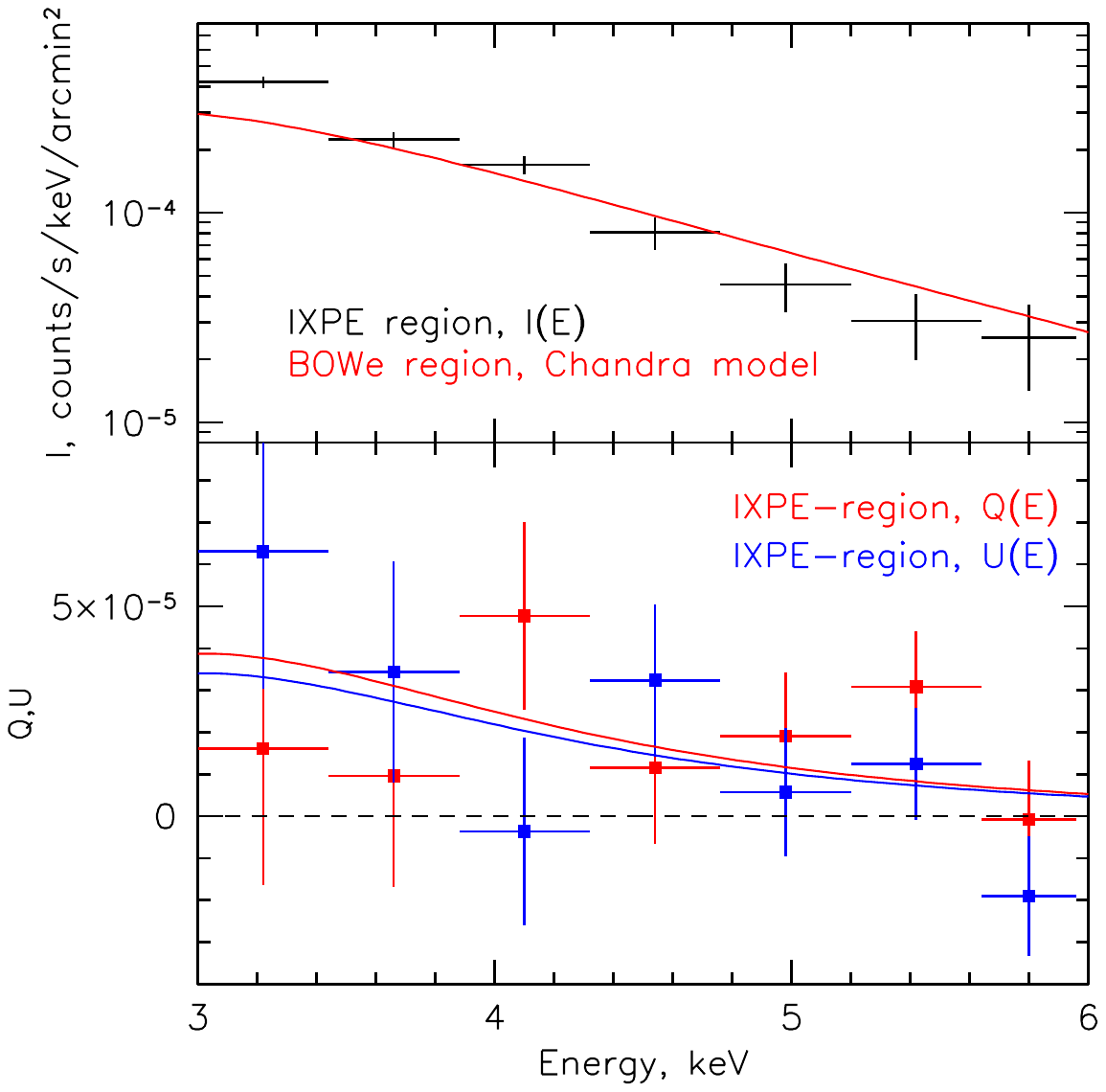}
\caption{IXPE spectra. {\bf Top panel:} IXPE total background-subtracted Stokes $I(E)$ spectrum in the 3-6 keV band from the IXPE region (see Fig.~\ref{fig:ximages} and \ref{fig:chandra_regs}) is shown with the black points. The red line in the top panel shows \textit{Chandra}'s Wings-e region model (see Fig.\ref{fig:spec_chandra}) convolved with the IXPE response. {\bf Bottom panel:} Stokes $Q(E)$ and $U(E)$ spectra of the IXPE region: red and blue points, respectively. The solid lines show the expected spectra for \textit{Chandra}'s Wings-e model and the best-fitting polarization degree ${\rm PD=}57\pm18$\% and angle ${\rm PA=}21\pm9$ degrees.
}
\label{fig:ixpe_spec}
\end{figure}

The parameters of the model are given in Table~\ref{tab:spec}. The central bright source, presumably the pulsar itself and possibly a compact part of its PWN, has the hardest spectrum, $\Gamma\sim 1.35$, while the Wings-n and Wings-e regions have significantly softer spectra, $\Gamma\sim 2.2-2.4$. Broadly, these spectral parameters are consistent with those obtained by  \cite{2002ApJ...568L.121Y} and \cite{2002ApJ...581.1148W}, given the difference in the choice of extraction regions and possible variability of the diffuse (reflected) emission. The IXPE region is clearly contaminated by thermal emission below 3-4 keV. However, already at $\sim 5\,{\rm keV}$, the fluxes from the Wings-e and IXPE regions are comparable. The Wings-e spectrum itself contains a non-negligible fraction of thermal emission. To set a lower limit on the degree of polarization, we conservatively assumed that the total intensity (i.e., Stokes $I(E)$) of the nonthermal component is described by the Wings-e spectrum (the red lines in Fig.~\ref{fig:spec_chandra}). We therefore model the IXPE's $Q(E)$ and $U(E)$ spectra assuming that $I(E)$ is known. With this approach, there are only two free parameters: the degree of polarization, $P,$ and the polarization direction, $\phi$. When fitting $Q(E)$ and $U(E)$ spectra, these parameters enter as pre-factors $P\cos 2\phi$ and $P\sin 2\phi$ in front of the \textit{Chandra} $I(E)$ model convolved with the IXPE response to polarized emission. Fig.~\ref{fig:ixpe_spec} shows the $I$,$Q$, and $U$ spectra obtained by IXPE, along with \textit{Chandra}'s Wings-e model convolved with the IXPE response. The IXPE response was averaged over three IXPE modules.

%IXPE: angle $31\pm 11$, pol.degree $0.48\pm 0.18$
%IXPE: 
%   7    5   cos2f      fi                  31.1695      +/-  10.5630      
%  17    4   constant   factor              0.514571     +/-  0.189727     

%------------------------
%-------------------------

\begin{figure}
\centering
\includegraphics[angle=0,clip=true,trim=1.5cm 14cm 1.7cm 3cm,width=0.99\columnwidth]{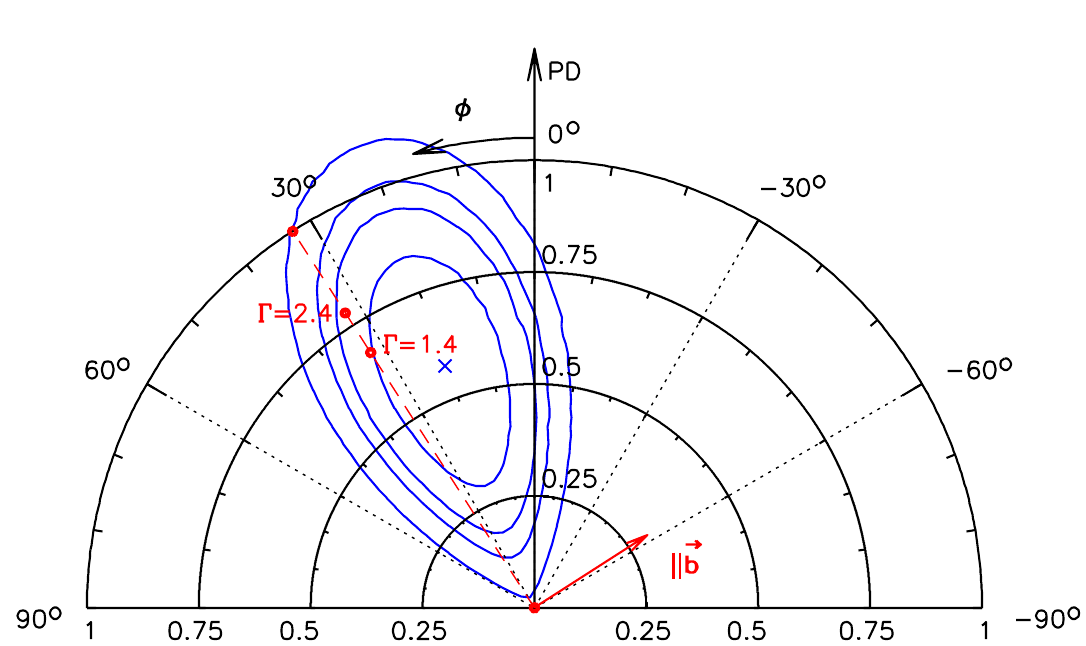}
\caption{Confidence regions for polarization degree (PD) and polarization angle (PA; equatorial coordinate system) derived from the IXPE data, assuming that the polarized emission corresponds to the Wings-e spectral model. The blue line shows the contours of the $\chi^2$ statistic after subtraction of the minimum value $\chi^2_{\rm min}$ reached at PD$\approx$0.57 and PA$\approx21^\circ$ (marked with a blue cross). The contours show $\Delta\chi^2=$2.3, 4.61, 6.17, and 9.21 levels, corresponding to 68.3\%(1$\sigma$), 90\%, 95.4\% (2$\sigma$), and 99\% confidence levels, respectively.
The red dashed line illustrates the expected PA in the case, when the electric field vector is parallel to the Galactic plane, and two circles mark the maximum polarization for synchrotron emission for $\Gamma=1.4$ and 2.4 (as labeled next to them). The outmost contour extends to PD values larger than 1. This is a plausible outcome of our procedure when the $Q(E)$ and $U(E)$ data are taken from  IXPE, while $I(E)$ is based on the \textit{Chandra} data extracted from a smaller region, which is supposed to be polarized and might not coincide with true polarized $I(E)$ seen by  IXPE.
}
\label{fig:chi2}
\end{figure}

The above assumptions yield the following best-fitting values:  ${\rm PD=}57\pm18$\% and angle ${\rm PA=}21\pm9$ degrees. Setting ${\rm PD=0}$ increases the value of $\chi^2$ by 10.37, which for two degrees of freedom implies a probability $\sim 5.3\times 10^{-3}$ of a random fluctuation. The corresponding confidence contours are shown in Fig.~\ref{fig:chi2}. The high value of the PD is a strong argument in favor of the synchrotron origin of the X-ray emission. The constraints on the polarization angle are shown in Fig.\ref{fig:outlook}. Within uncertainties, the polarization plane is perpendicular to the bright X-ray structure and, also, to the direction of radio filaments co-spatial with the extended X-ray source. This suggests that both the radio and X-ray emissions are due to the synchrotron radiation of relativistic electrons in the same magnetic structures.

We note here that when calculating the degree of polarization, we conservatively assumed that essentially all nonthermal flux coming from the IXPE region is polarized (i.e., model Wings-e). If instead only the emission coming from the Wings-n region or from the PWN itself is polarized, then the degree of polarization will only be higher.
%so that the formal estimate of PD as $\sqrt{Q^2+U^2}/I$ can be easily larger than one. 
For instance, substituting the \textit{Chandra}-based Wings-e model with the Wings-n model for $I(E)$ yields  PD=$160\pm60$\%. A high degree of polarization ($> 100$\%) is a plausible outcome of using the incorrect assumption of what fraction of the total signal comes from the polarized source.   Apart from the statistical errors already included in the uncertainties of PD estimates, there might be additional uncertainties associated with the limited astrometric quality of IXPE data, although our choice of a large region for spectral extraction will have minimized these uncertainties.

The choice of the background region might affect the shape and normalization of the Wings-e spectral model derived from the \textit{\textit{Chandra}} spectra. Both the variations in the diffuse emission level and in the absorbing column density can contribute. Experiments with four different background regions have shown that the predicted flux in the 3-6~keV band does not vary by more than 20\%. One can therefore consider this as an additional source of systematic uncertainties in the measured $PD$ value, which, however, does not affect the significance of the polarization detection. This background-induced uncertainty is subdominant to the uncertainties associated with the choice of the Wings-e region to extract the reference $I(E)$ spectrum. 
We also note that the presence of the bright and time-variable intermediate polar (IP) CXOUGC J174622.7–285218  \citep{2009ApJS..181..110M} in close vicinity of G0.13-0.11 (see Fig.~\ref{fig:chandra_regs}) can be considered as a possible source of the polarized signal contamination for IXPE. However, as is evident from Fig.~\ref{fig:spec_chandra}, its flux at $\sim 4\,{\rm keV}$ is similar to that of the Wings-n region. This means that if this IP is the dominant source of polarization, its emission should be almost 100\% polarized, which is unlikely. For example, \cite{2004A&A...423..495M} estimated the maximum degree of polarization in magnetic cataclysmic variables of about 4\% produced by the scattering of the thermal radiation.

We reiterate here that spatially resolving contributions of the IP and the diffuse emission can be easily done in the \textit{Chandra} data. However, for IXPE, the coarser angular resolution and the uncertainties with the absolute astrometry (i.e., an apparent shift) complicate the procedure. The photons from IP and the diffuse emission associated with the PWN are mixed in the IXPE images.  The choice is therefore between two options: i) make a smaller and shifted region that excludes IP using \textit{Chandra} images, but might miss a hard--to-quantify fraction of the diffuse/polarized emission or ii) make a larger region (based on IXPE images) that captures most of the diffuse emission but is contaminated by the IP contribution, which can be removed from the PWN spectrum using \textit{Chandra} data.  We decided that the latter option is a more transparent approach, which, however, requires an assumption that the IP emission is not strongly polarized, as supported by the IP polarization models of \cite{2004A&A...423..495M}.

%------------------------
%-------------------------

\section{Discussion and conclusions}
\textit{\emph{The} IXPE} data provide two main observational results: (i) the strong polarization of G0.13-11 X-ray emission, ${\rm PD=}57\pm18$\%, and (ii) the polarization direction almost along the Galactic plane. This directly proves that X-ray emission is of synchrotron origin and that the orientation of the magnetic field follows a subsystem of the Radio Arc NTFs\footnote{The large Faraday rotation measure ${\rm RM} \sim {\rm a \, few} \, 10^3\, {\rm rad \, m^{-2}}$ \citep[e.g.,][]{2019ApJ...884..170P,2021ApJ...923...82P} complicates the PA determination at radio frequencies and limits reliable polarization measurements to patches of the brightest radio filaments. In contrast, polarization in the X-ray regime does not suffer from Faraday rotation.}, which are elongated perpendicularly to the Galactic plane. These statements are essentially model independent. 

These data also prove that the field is well ordered.  Indeed, for a perfectly aligned magnetic field and a power-law distribution of relativistic leptons over energy  $\propto \gamma^{-p}$ with index $p$, the degree of the synchrotron emission polarisation is \citep[e.g.,][]{2011hea..book.....L}
\begin{align}
{\rm PD}_{\rm max}=\frac{3p+3}{3p+7}=\frac{3\Gamma}{3\Gamma+2},
\end{align}
where $\Gamma=(p+1)/2$ is the photon index of the observed spectrum (for $\Gamma$ between 1.4 and 2.2, the maximum degree of polarization is between 68 and 78\%). 
%While this dependence can in principle be used to infer independent constraints on the slope (a lower limit) of the leptons' spectrum in combination with a degree of the field ordering, in practice the statistical and systematic uncertainties are \nb{quite} large for this exercise. 
Using the formulation for depolarization of synchrotron emission in a turbulent magnetic field by \citet{Bandiera_Petruk16a}, we estimated a ratio of turbulent versus ordered magnetic field energy to be likely smaller than $\sim 1.5$.
%We, therefore, only state that 
However, the observed PD$=57\pm18$\% is consistent (within uncertainties) even with a perfectly ordered field (see Fig.~\ref{fig:chi2}). We note that a high level of polarization can be achieved even in a fully turbulent regime if the turbulence is anisotropic (typically at a level of 2--4), given that polarization is sensitive to the alignment of the magnetic field with a certain direction
and is insensitive to field reversals \citep{Bandiera_Petruk16a}. This preferred direction should be maintained across the IXPE spectral extraction region. Indeed, current models of X-ray emission in bright jet-like features associated with PSRs advocate the amplification of the magnetic field via some form of streaming instability,  to justify magnetic field values that are typically inferred to be an order of magnitude higher than the surrounding ISM \citep{Bandiera08a}. This might not be needed in the GC environment, where the ambient magnetic field can be high.

The X-ray emission of G0.13-0.11 is co-spatial with a Fermi source 4FGL J1746.4-2852 \citep{2020ApJS..247...33A} at GeV energies and a TeV-source HESS J1746–285 \citep{2018A&A...612A...9H}. The standard leptonic model for broadband radio-TeV objects usually relies on two processes: the synchrotron and inverse Compton (IC) emissions. We defer an in-depth discussion for a forthcoming publication, but we note that the following set of fiducial parameters might explain some of the main characteristics of the G0.13-0.11 broad-band spectrum: $B\sim 100\,\mu{\rm G}$ is relevant for the radio--X-ray regime (synchrotron emission) and a combination of the CMB plus local $\sim$eV radiation field with the energy density of the order of $100\,{\rm eV\,cm^{-3}}$ \citep[e.g.,][]{2017MNRAS.470.2539P,2019APh...107....1N} is relevant for the photons in the GeV-TeV range produced via IC scatterings. Such a value of the magnetic field strength agrees with the equipartition arguments \citep[e.g.,][]{2005AN....326..414B} based on the measured X-ray flux and the broad-band photon index $\Gamma=2$, although allowing the index to vary between 1.6 and 2.4 changes $B$ between $\sim 20$ and $\sim 800\,{\rm \mu G}$, respectively \citep[see, e.g.,][for discussion oon various arguments in favor of weaker and stronger fields]{2006JPhCS..54....1M}.  The adopted value of $B\sim 100\,\mu{\rm G}$ implies that the field energy density is of the same order as the energy density of the radiation and, therefore, the lifetime of leptons emitting at a given frequency is close to the maximum. In this case, X-rays are produced by the leptons with $\gamma\sim 3\times 10^{7}$ which have a lifetime consistent with the size of G0.13-0.11, provided that they can propagate with a sizable fraction ($\sim 0.2$) of the speed of light. For photons with TeV energies, produced by IC scattering of $\sim$eV photons, the cross-section is already subject to the Klein-Nishina suppression, while for the CMB photons, the scattering is still in the Thomson regime, where the cross-section is maximal.

In connection with the fast propagation of relativistic electrons along a filament with a rather strong magnetic field, we note an interesting possible similarity with galaxy clusters, where extended structures in the radio bands are observed both in the cores and outskirts \citep[e.g.,][]{2021NatAs...5.1261B,2022ApJ...935..168R}. These structures often have very similar spectral properties \citep[e.g.,][]{2020A&A...636A..30R}. This suggests that relativistic electrons can propagate along the filament with velocities significantly larger than the sound or Alfven speeds in the bulk of the thermal intracluster medium  \citep[][]{2023A&A...670A.156C}. The filaments in both classes of objects might be threads of a strong magnetic field embedded in the medium with a weaker magnetic field. This does not exclude possible amplification of the magnetic field by streaming particles from the PSR.

Apart from the polarized synchrotron X-ray emission of G0.13-0.11, there are many more radio and (fainter) X-ray threads in the same region that are visible in \textit{Chandra} data. Presumably, they are polarized too and, therefore, can contribute to the overall polarization signal from the GC region. It is also possible that even fainter threads are present that are too weak to be detected individually. Since the direction of polarization might be set by the ambient magnetic field, this background polarization signal will also be polarized along the Galactic plane (at least in the areas where "vertical" NTFs are present).

Yet another source of polarized emission is the Compton-scattered emission of the Sgr A* flare(s) that happened hundreds of years ago. This scenario, motivated by the observed hard X-ray continuum and fluorescent lines of neutral iron from dense molecular clouds, was put forward in the 90s \citep{1993ApJ...407..606S,1993Natur.364...40M,1996PASJ...48..249K}. If the continuum emission is scattered light from Sgr~A*, it has to be polarized \citep{2002MNRAS.330..817C}  perpendicularly to the direction towards Sgr~A*. A weak polarized emission from the area within a few arcminutes of G0.13-0.11 was indeed detected in the first IXPE observations of the GC region \citep{2023Natur.619...41M}. The emission from G0.13-0.11 itself was excluded from that analysis to avoid potential contamination. Now we see that there was a good reason to do so. Somewhat ironically, the polarization angles from X-ray threads and reflection are approximately orthogonal to each other and, therefore, can attenuate the net polarization signal if combined. The presence of many faint polarized threads is, therefore, a complication (an interesting one!) for the analysis of IXPE data from the GC. 
The limited astrometric accuracy of IXPE does not allow for a clean spatial separation of these two components, but a combination of IXPE data with the high spatial resolution of \textit{Chandra} and \textit{XMM-Newton} and \textit{Chandra} spectra can be used to constrain their contributions. This will be reported elsewhere.

We conclude by saying that IXPE observations show that X-ray emission from G0.13-0.11 is strongly polarized ($\sim 60\%$) in the direction perpendicular to the bright wings-like X-ray structure and to the nearby radio-emitting filaments, which are part of the GC Radio Arc. While the IXPE angular resolution is not sufficient to spatially resolve polarized emission into individual subarcminute components (i.e., the wings and the fainter diffuse emission to the east of them), the statement of the high degree of polarization is robust.
These    
measurements prove that X-ray emission of filaments associated with BSPWNs has a synchrotron origin, as expected in the baseline scenario of these objects. In particular,  X-ray threads of 
G0.13-0.11 are produced by synchrotron emission of $\sim 10$~TeV leptons gyrating in the ordered magnetic field perpendicular to the Galactic plane. This finding corroborates the scenario that PWNs power not only the X-ray emission but also radio emission of filaments in the GC region \citep[e.g.,][]{2002ApJ...581.1148W,2017SSRv..207..235B,2019MNRAS.489L..28B}. 
To this end, a particularly interesting question (to be addressed elsewhere) would be whether the entire Radio Arc is powered by the same mechanism (see \citealt{2002ApJ...581.1148W} for a relevant discussion).

\vspace{0.3cm}

\section*{Acknowledgements}

The Imaging X-ray Polarimetry Explorer (IXPE) is a joint US and Italian mission.  The US contribution is supported by the National Aeronautics and Space Administration (NASA) and led and managed by its Marshall Space Flight Center (MSFC), with industry partner Ball Aerospace (contract NNM15AA18C).  The Italian contribution is supported by the Italian Space Agency (Agenzia Spaziale Italiana, ASI) through contract ASI-OHBI-2022-13-I.0, agreements ASI-INAF-2022-19-HH.0 and ASI-INFN-2017.13-H0, and its Space Science Data Center (SSDC) with agreements ASI-INAF-2022-14-HH.0 and ASI-INFN 2021-43-HH.0, and by the Istituto Nazionale di Astrofisica (INAF) and the Istituto Nazionale di Fisica Nucleare (INFN) in Italy.  This research used data products provided by the IXPE Team (MSFC, SSDC, INAF, and INFN) and distributed with additional software tools by the High-Energy Astrophysics Science Archive Research Center (HEASARC), at NASA Goddard Space Flight Center (GSFC).

IK acknowledges support by the COMPLEX project from the European Research Council (ERC) under the European Union’s Horizon 2020 research and innovation program grant agreement ERC-2019-AdG 882679.
RK, AV, and WF acknowledge support from the Smithsonian Institution, the High Resolution Camera Project through NASA contract NAS8-03060, and \textit{Chandra} grant GO1-22136X.  NB is supported by the INAF MiniGrant "PWNnumpol - Numerical Studies of Pulsar Wind Nebulae in The Light of IXPE". E.Co., A.D.M., R.F., P.So., S.F., F.L.M., F.Mu. are partially supported by MAECI with grant CN24GR08 “GRBAXP: Guangxi-Rome Bilateral Agreement for X-ray Polarimetry in Astrophysics”.

The MeerKAT telescope is operated by the South African Radio Astronomy Observatory, which is a facility of the National Research Foundation, an agency of the Department of Science and Innovation.

\bibliographystyle{mnras}
\bibliography{ref.bib} % if your bibtex file is called example.bib
%\bibliography{ref} % if your bibtex file is called example.bib

%%%%%%%%%%%%%%%%%%%%%%%%%%%%%%%%%%%%%%%%%%%%%%%%%%

%%%%%%%%%%%%%%%%%%%% REFERENCES %%%%%%%%%%%%%%%%%%

% The best way to enter references is to use BibTeX:

%\bibliographystyle{mnras}
%\bibliography{example} % if your bibtex file is called example.bib

%%%%%%%%%%%%%%%%%%%%%%%%%%%%%%%%%%%%%%%%%%%%%%%%%%

%%%%%%%%%%%%%%%%% APPENDICES %%%%%%%%%%%%%%%%%%%%%

%\appendix

%\section{Some extra material}

%%%%%%%%%%%%%%%%%%%%%%%%%%%%%%%%%%%%%%%%%%%%%%%%%%

% Don't change these lines
%\bsp   % typesetting comment
\label{lastpage}
\end{document}